\def\be{\begin{equation}}
\def\ee{\end{equation}}
\def\ba{\begin{array}}
\def\ea{\end{array}}

\documentclass[11pt]{article}
\usepackage{amsfonts}
\usepackage{amssymb}
\usepackage{dsfont}
\usepackage{amsmath,amssymb,amstext,amsfonts,array,hhline,tabularx,graphics}
\usepackage{amsthm}

\usepackage{latexsym}
\usepackage{arydshln}
\usepackage{epsfig,graphicx, color}
\usepackage{amsmath}
 \usepackage{cite}
\usepackage{amsthm}

\newtheorem{thm}{\bf Theorem}

\newtheorem{prop}{\bf Proposition}

\theoremstyle{plain}
\begin{document}
\parskip=3pt
\parindent=18pt
\baselineskip=20pt \setcounter{page}{1}

 \title{\large\bf A unifying separability criterion based on extended correlation tensor}
\date{}

\author{Xiaofen Huang$^{1, \ast}$ , Tinggui Zhang$^{1}$,  Naihuan Jing$^{2, 3}$ \\[10pt]
\footnotesize
\small 1 School of Mathematics and Statistics, Hainan Normal University, Haikou 571158, China\\
\small 2 Department of Mathematics, North Carolina State University,
Raleigh, NC27695, USA\\
\small 3 Department of Mathematics, Shanghai University, Shanghai 200444, China}
\date{}

\maketitle

\centerline{$^\ast$ Correspondence to  huangxf1206@163.com  }
\bigskip

\begin{abstract}
Entanglement is fundamental inasmuch because it
rephrases the quest for the classical-quantum demarcation
line, and it also has potentially enormous practical
applications in modern information technology.
In this work, employing the approach of matrix decomposition, we introduce and formulate a practicable criterion for separability based on the correlation tensor. It is interesting that this criterion unifies several relevant separability criteria proposed before, even stronger than some of them.
Theoretical analysis and detailed examples demonstrate its availability and feasibility for entanglement detection. Furthermore we build a family of entanglement witnesses using the criterion according to its linearity in the density operator space.
\end{abstract}


\section{Introduction}

Quantum entanglement is an intrinsic feature of composite quantum systems and is
rooted in the superposition principle of quantum mechanics, which is widely recognized as a significant practical resource for quantum information processing, like quantum teleportation \cite{telepor}, quantum algorithms \cite{algori1, algori2} and quantum cryptography \cite{cryp}, that will enable a ``quantum computer'' to outperform classical devices, in this sense, it is possible to herald a new technological revolution.

Both theoretically and experimentally,  considerable efforts have been made to understand entanglement hidden in compound systems consisting of distinguishable quantum states.  However, there are still many challenging issues remain unsolved,  one of the key problems is to answer in full generality whether a given quantum state is entangled or not.
Mathematically, a bipartite quantum state $\rho$  acting on  the compound quantum system $\mathcal{H}_A\otimes \mathcal{H}_B$ is called separable, if it has a convex representation of product states
\begin{equation}
\rho=\sum_{i}p_i\rho_{i}^{A}\otimes \rho_i^{B},
\end{equation}
where $p_i$ is a probability distribution,  $\rho_{i}^{A}$  and $ \rho_i^{B}$ are the density matrices of subsystems $\mathcal{H}_A$ and $\mathcal{H}_B$ respectively,
otherwise $\rho$ is entangled.

In recent decades, a number of meaningful approaches to detect entanglement have
been proposed. In case of full information for density matrix, the celebrated PPT criterion \cite{PPT, Hordek} is a strong operable and  feasible approach for entanglement detection,  which claims that a state is separable if and only if it is a positive partial transpose for low dimensional systems. However, there are some entangled states escaped from the PPT criterion in higher-dimensional systems. As a generalization of the PPT criterion, entanglement monotonicity is put forward  in terms of moments of the partially transposed density matrix \cite{moment, Ali2023, Neven2021}. The advantage of the PT-moments is that they can be experimentally measured through global random unitary matrices \cite{random1} or local randomized measurements \cite{random2} based on quantum shadow estimation \cite{shadow}.

Essentially, the partial transpose matrix $\rho^{\tau}$ endowed with elements $(\rho^{\tau})_{ij, kl}=(\rho)_{il, kj}$ is from the partial transpose transformation with respect to subsystem $B$. Permutating the indicators of $\rho_{ij, kl}$, we can obtain the realigned matrix $\rho^{R}$ with elements $(\rho^{R})_{ij, kl}=(\rho)_{ik, jl}$. Realignment criteria\cite{reag} is another vital approach for detecting entanglement.
Besides the PT-moments criterion, there are also many other criteria based on moments such as the one based on rearrangement moments \cite{zhang, Agg2023, Agg2024}
and the $\Lambda$-moment criterion \cite{moment1} proposed recently.

Furthermore, there are additional criteria \cite{oth1, oth2, Jin2023} that are simple for applications
and in particular allow us to distinguish many PPT entangled states. The prominent examples are the correlation matrix (CM) and computable cross norm  (CCNR) criteria \cite{vicente2007,ccnr} based on the correlation tensor. The CCNR criterion exhibits
a powerful capability in detecting PPT entanglement. It has been reformulated in
several forms and also extended to multipartite systems \cite{rud}. The correlation tensor reveals insights on nonlocality of the quantum system, thus it helps characterizing and quantifying entanglement. Separability criteria based on correlation tensors were also analyzed in Ref. \cite{liming2014, shen2016, chang2018, Sarbicki2020, qiao2022, vicente2011PRA, correlation2016, JinZh22, Zhao23} and so on.
Experimental detection of entanglement is facilitated by the entanglement witness that is expressible as a linear Hermitian operator with at least one negative eigenvalue \cite{witness1, witness2, witness3, witness4},  that has a nonnegative mean value for separable states, and that carries a negative mean value for entangled states. The universal approach involving witnesses provides a necessary and sufficient condition to detect entanglement. Entanglement witness can also be used to characterize quantum nonlocality, and it has emerged as an effective method to
study quantum state discrimination \cite{dis1, dis2}.

In this work we present a computable separability criterion based on the Bloch representation of density matrix, and analyze its capability of detecting entanglement. By
constructing an extended correlation tensor via orthogonal bases of subsystems, our separability criteria turns out in alignment with several existing
approaches such as de Vicente's criterion \cite{vicente2007}, CCNR \cite{ccnr} and Li's criterion \cite{liming2014}, thus it appears to be
an unifying method to detect entanglement. A detailed example illustrates its superiority in analyzing entanglement. Moreover, this approach also
provides a family of entanglement witnesses due to the linearity of the trace function.

\section{Separability criterion for bipartite systems  }
Let us begin by recalling some basic definitions and notations.
Let $\{G_{i}\}$ denote a fixed basis  of orthogonal operators in the space $\mathcal{L}(\mathcal{H})$ endowed with orthogonal relation
$(G_{i}, G_{j})=\kappa \delta_{ij}$,  where $\kappa$ is a positive number no less than one, and the
Hilbert-Schmidt inner product is defined as usual: $( X, Y): = \rm{Tr}( X^{\dag}Y)$.
Any state $\rho$ acting on the quantum system $\mathcal{H}_d$ with dimensional $d$ can be rewritten relative to the basis $\{G_{i}\}$ in the form
\begin{equation}
\rho=\frac{1}{d}I+\sum_i\nu_iG_i.
\end{equation}
The Bloch vector $ \textit{\textbf{v}}: =(\nu_i)$ lies in the space $\mathbb{C}^{d^2-1}$, and the trace condition $\mathrm{Tr} \rho^2\leq 1$ implies that the length of the Bloch vector fulfills the inequality $\|\textit{\textbf{v}}\|^2\leq\frac{d-1}{\kappa d}$.

 Consider further a bipartite quantum system $\mathcal{H}_{d_A} \otimes \mathcal{H}_{d_B}$ with the subsystems of dimensions $d_A$ and $d_B$ respectively.
  Choose an arbitrary traceless orthogonal basis $ \{G_{i}^{A}\}$ in the operator space $\mathcal{L}(\mathcal{H}_{A} )$ such that
  $(G_{i}^{A}, G_{j}^{A})=\kappa_A \delta_{ij}$, and similarly $ \{G_{j}^{B}\}$ is a likewise basis of the space $\mathcal{L}(\mathcal{H}_{B} )$
  with $(G_{i}^{B}, G_{j}^{B})=\kappa_B \delta_{ij}$. For simplicity we assume that $G_0^A=\sqrt{\frac{\kappa_A}{d_A}}I_A$ and $G_0^B=\sqrt{\frac{\kappa_B}{d_B}}I_B$. It is worth noting that the operator bases mentioned above should be traceless but not essential for hermiticity, known as the unitary Heisenberg-Weyl operators basis \cite{huang2022} 
 and the polarization operators basis \cite{reinhold}.

For a given bipartite state $\rho$ in $\mathcal{L}(\mathcal{H})_{d_A} \otimes \mathcal{L}(\mathcal{H})_{d_B}$, one can decompose it in terms of  the orthogonal bases $\{G_{i}^A\}$ and $\{G_{j}^B\}$.
Namely,  the generalized Bloch representation of $\rho$ can be represented as
\begin{equation}\label{bloch}
\rho=\frac{I_A}{d_A}\otimes \frac{I_B}{d_B}+\sum_{i=1}^{d_A^2-1} r_i G_i^A \otimes \frac{I_B}{d_B}+\sum_{j=1}^{d_B^2-1} s_j\frac{I_A}{d_A}\otimes G_j^B
+\sum_{i=1}^{d_A^2-1}\sum_{j=1}^{d_B^2-1}t_{ij}G_i^A\otimes G_j^B.
\end{equation}
where the coefficients can be expressed as $r_i=\frac{1}{\kappa_A}\mathrm{Tr} \rho G_i^A\otimes I_B$, $s_j=\frac{1}{\kappa_B}\mathrm{Tr} \rho I_B\otimes G_j^B$ and $t_{ij}=\mathrm{Tr} \rho G_i^A\otimes G_j^B$. The correlation matrix or correlation tensor is defined by the matrix $T: =(t_{ij})$ of size $(d_A^2-1)\times (d_B^2-1)$ wherein quantum information is stored,
which is widely used in detecting entanglement \cite{liming2014, shen2016, correlation2016}.

  To elaborate our separability criterion, we construct an extended correlation tensor in the form of a second order block matrix.
  For any nonnegative numbers $x$, $y$ and positive integer $n$, the extended correlation tensor of $\rho$ can be formed as
\begin{equation}
\mathcal{M}_{x, y}^{(n)}:=\left(\begin{array}{cc}
                                     \frac{xy}{\sqrt{\kappa_A\kappa_Bd_Ad_B}}  E_{n\times n} & \frac{x}{\sqrt{\kappa_A d_A}} \omega_n(\textbf{s})^t \\
                                      \frac{y}{\sqrt{ \kappa_B d_B}} \omega_n(\textbf{r}) & T\\
                                     \end{array}\right),
\end{equation}
where $E_{k\times l}$ is the $k\times l$ matrix with all entries being $1$. The superscript $t$ stands for transpose, and for any column vector $\textit{\textbf{u}}$, $\omega_n(\textit{\textbf{u}})$ is the rank one matrix made of the same column vector $\textit{\textbf{u}}$ given by
$$
 \omega_n(\textit{\textbf{u}}):=\underbrace{(\textit{\textbf{u}}, \cdots, \textit{\textbf{u}} )}_{\text{n columns}}.
$$

It should be noted that the extended correlation tensor $\mathcal{M}_{x, y}^{(n)}$ consisting of entries $m_{ij}$ with size $(d_A^2+n-1)\times (d_B^2+n-1) $, thus it is different from the usual
augmented correlation tensor $\tilde{T}=\begin{pmatrix} 1 & \textbf{s}^t\\
\textbf{r} & T\end{pmatrix} $ of size $d_A^2\times d_B^2$. Of course it owns close relations with  the coefficients in \eqref{bloch}, that is,
\begin{equation}\label{relation}
m_{ij}=\left\{
                     \begin{array}{ll}
                       \frac{xy}{\sqrt{\kappa_A\kappa_Bd_Ad_B}} & \hbox{$1\leq i\leq n, ~~1\leq j\leq n$;} \\
                       \frac{x}{\sqrt{\kappa_A d_A}}s_{j-n}& \hbox{$1\leq i\leq n,~~ n<j\leq d_B^2+n-1$;} \\
                       \frac{y}{\sqrt{ \kappa_B d_B}}r_{i-n}& \hbox{$n< i\leq d_A^2+n-1, ~~1\leq j\leq n$;} \\
                       t_{i-n, j-n} & \hbox{$n< i\leq d_A^2+n-1,~~ n<j\leq d_B^2+n-1$.}
                     \end{array}
                   \right.
\end{equation}
This implies that every element $m_{ij}$ is in the form of the mean value of $\rho$ and the operator basis of the compound quantum system. It is obvious that extended correlation tensor is a generalization of canonical correlation tensor. In case of $n=0$ or $x=y=0$, these two matrices coincide with each other. When $n=1$, it reduces to the augmented
correlation tensor $\tilde{T}$ with appropriate choice of $x, y$.

Recall the trace norm $\|M\|_{tr}$ of matrix $M$, also called the Ky Fan norm in some literatures, is defined by
the sum of singular values of $M$.  We use this to derive a separability criterion based on extended correlation tensors as follows.

\begin{thm}\label{thmbi1}
For a separable bipartite state $\rho$ in space $\mathcal{L}(\mathcal{H})_{d_A} \otimes \mathcal{L}(\mathcal{H})_{d_B}$, the extended correlation tensor of $\rho$  should satisfy the following inequality:
\begin{equation}\label{mycri}
\| \mathcal{M}_{x, y}^{(n)} \|_{tr}\leq \sqrt{(\frac{nx^2+d_A-1}{ \kappa_Ad_A})(\frac{ny^2+d_B-1}{\kappa_Bd_B})},
\end{equation}
where $x$, $y$ are nonnegative and $n$ is a positive integer.
\end{thm}

\textbf{Proof}: Suppose $\rho$ is separable, then it can be represented as a convex combination of product states, i. e.,  $\rho=\sum_ip_i|\varphi_i\rangle_A\langle\varphi_i|\otimes |\psi_i\rangle_B\langle\psi_i|$, where $p_i$ is a probability distribution,  $|\varphi_i\rangle_A$ and $|\psi_i\rangle_B$ are the pure states in $\mathcal{H}_A$ and $\mathcal{H}_B$, respectively. This implies there exists
 vectors $\textbf{u}_i\in \mathbb{C}^{d_A^2-1}$, $\textbf{v}_j\in \mathbb{C}^{d_B^2-1}$ such that
\begin{equation}
T=\sum_i p_i\textbf{u}_i\textbf{v}_i^t,~~~ \textbf{r}=\sum_i p_i\textbf{u}_i,~~~ \textbf{s}=\sum_i p_i\textbf{v}_i,
\end{equation}
therein $\textbf{u}_i$ and $\textbf{v}_i$ are the Bloch vectors in subsystems  $\mathcal{H}_{d_A}$ and $\mathcal{H}_{d_B}$ respectively, and the lengths of them should satisfy the following conditions,
\begin{equation}\label{length}
\|\textbf{u}_i \|^2=  \frac{d_A-1}{ \kappa_A d_A},~~~~ \|\textbf{v}_i \|^2= \frac{d_B-1}{\kappa_B d_B}.
\end{equation}
It also signifies that the extended correlation tensor  $\mathcal{M}_{x, y}^{(n)}$ can be rewritten as a convex combination as below
\begin{equation}
 \mathcal{M}_{x, y}^{(n)}=\sum_i p_i\left(\begin{array}{cc}
                                     \frac{xy}{\sqrt{\kappa_A\kappa_Bd_Ad_B}}  E_{n\times n} & \frac{x}{\sqrt{\kappa_A d_A}} \omega_n(\textbf{s})^t \\
                                      \frac{y}{\sqrt{\kappa_B d_B}} \omega_n(\textbf{r}) & \textbf{u}_i\textbf{v}_i^t\\
                                     \end{array}\right).
\end{equation}

Now we introduce two alternative vectors $\overline{\textbf{u}}_i=\left(\begin{array}{c} \frac{x}{\sqrt{\kappa_A d_A}} E_{n\times 1} \\  \textbf{u}_i \\ \end{array} \right)$ and $\overline{\textbf{v}}_i=\left(\begin{array}{c}\frac{y}{\sqrt{\kappa_B d_B}}  E_{n\times 1} \\  \textbf{v}_i \\ \end{array} \right)$, using decomposition method of matrix one can obtain
\begin{equation}
\mathcal{M}_{x, y}^{(n)}
   = \sum_i p_i
   \left(\begin{array}{c}\frac{x}{\sqrt{\kappa_A d_A}} E_{n\times 1} \\\textbf{u}_i \\\end{array}\right)
   \left( \begin{array}{cc}\frac{y}{\sqrt{\kappa_B d_B}} E_{1\times n} & \textbf{v}_i^t \\ \end{array}\right)
  =\sum_i p_i \overline{\textbf{u}}_i \overline{\textbf{v}}_i^t.
\end{equation}

 It follows from \eqref{length} that the extended tensor $\mathcal{M}_{x, y}^{(n)}$ can be reformulated as follows:
\begin{equation}
\begin{aligned}
\| \mathcal{M}_{x, y}^{(n)} \|_{tr} &\leq  \sum p_i \|\overline{\textbf{u}}_i \overline{\textbf{v}}_i^t \|_{tr}= \sum p_i \|\overline{\textbf{u}}_i \|\|\overline{\textbf{v}}_i \| \\
   &= \sqrt{(\frac{nx^2+d_A-1}{ \kappa_Ad_A})(\frac{ny^2+d_B-1}{\kappa_Bd_B})}.
\end{aligned}
\end{equation}
This completes the proof. We remark that  the equality holds only for separable pure states.

Usually,  one can rewrite the Bloch representation of density matrix using a factor  as below
\begin{equation}
\rho=\frac{1}{d_Ad_B}(I_A\otimes I_B+\sum_{i=1}^{d_A^2-1}\widehat{r}_iG_i^A\otimes I_B+\sum_{j=1}^{d_B^2-1}\widehat{s}_jI_A\otimes G_j^B+\sum_{i=1}^{d_A^2-1}\sum_{j=1}^{d_B^2-1}\widehat{t}_{ij}G_i^A\otimes G_j^B).
\end{equation}
In this context the length of partial Bloch vector enjoys inequality $\|\textit{\textbf{v}}\|^2\leq \frac{d^2-d}{\kappa}$.  Analogously, one can define
 the corresponding extended correlation tensor as
\begin{equation}
\widehat{\mathcal{M}_{x, y}^{(n)}}:=\left(\begin{array}{cc}
                                    xy  E_{n\times n} & x \omega_n(\widehat{\textbf{s}})^t \\
                                     y \omega_n(\widehat{\textbf{r}}) &\widehat{ T}\\
                                     \end{array}\right).
\end{equation}

As a result, it leads to another criterion for bipartite separability based on the trace norm of the extended correlation tensor $\widehat{\mathcal{M}_{x, y}^{(n)}}$.
\begin{prop}\label{prop}
For a separable state $\rho$ in bipartite quantum system $\mathcal{L}(\mathcal{H})_{d_A} \otimes \mathcal{L}(\mathcal{H})_{d_B}$, the trace norm of extended correlation tensor is bounded
as follows:
\begin{equation}\label{mycri2}
\| \widehat{\mathcal{M}_{x, y}^{(n)}} \|_{tr}\leq \sqrt{(nx^2+\frac{d_A^2-d_A}{ \kappa_A})(ny^2+\frac{d_B^2-d_B}{\kappa_B})},
\end{equation}
where $x$, $y$ are nonnegative and $n$ is a positive integer.
\end{prop}
The proof of this Propsition is similar to that of Theorem \ref{thmbi1}, one just uses instead the square length of partial Bloch vector with $\frac{d^2-d}{\kappa}$ .

We remark that
the inequalities (\ref{mycri}) and (\ref{mycri2}) imply that there is an upper bound for the extended correlation tensor in separable states,
passing the threshold is only possible through entanglement. What is more, these bipartite separability criteria proposed here can be generalized to fully separability criteria for multipartite quantum states by the method of matricizations of extended correlation tensor \cite{shen2016, huang2024, lenny2023}.

\section{Relations to other separability criteria}

We will show that the extended correlation matrix can give us a better bound for an entangled quantum state.
The main theorem  proposes a necessary condition for separability of bipartite quantum state, it is feasible to detect entanglement once it violates the criterion inequality.
Moreover, one of advantages of the new criterion is to combine several separability criteria in a unified form. As we shall discuss in more detail below, the
choices of the orthonormal operator bases and the parameters $x, y, n$ can help improving the separability condition.


Recall the well-known computable cross-norm criterion (CCNR) \cite{ccnr} based on the correlation tensor gives the bound for the trace norm of $T$ for a separable state: $\parallel T \parallel_{tr}\leq 1$.
Taking parameters $x=y=0$ in the inequality \eqref{mycri}, one can get the following expression immediately
\begin{equation}
\parallel T \parallel_{tr}=\parallel \mathcal{M}_{x, y}^{(n)} \parallel_{tr}\leq \sqrt{(\frac{d_A-1}{\kappa_Ad_A})(\frac{d_B-1}{\kappa_Bd_B}})\leq 1.
\end{equation}
Therefore our criterion in Theorem \ref{thmbi1} reduces to CCNR in this special case.  Furthermore it provides a much stronger bound than CCNR.

Recently, Sarbicki \textit{et al}. proposed a family of separability criteria \cite{Sarbicki2020} based on the Bloch decomposition of density matrix, and
the criteria include several criteria known before as special cases. Furthermore, taking isotropic state as an example and its analysis illustrate that those criteria are equivalent to the enhance realignment criterion \cite{guo2008}  which is said to be the strongest effectively computable simplification of correlation matrix criterion \cite{sar2021}.
If we choose parameters $\kappa_A=\kappa_B=n=1$  for the criterion inequality (\ref{mycri}),
we also obtain the key criterion formulated in Ref.\cite{Sarbicki2020}:
\begin{equation}\label{Sar}
\|\mathcal{M}_{x, y}^{(n)}\|_{tr}\leq  \sqrt{(\frac{x^2+d_A-1}{ d_A})(\frac{y^2+d_B-1}{d_B})}.
\end{equation}
From this perspective, our criterion is also equivalent to both enhance realignment criterion and separability criteria given in \cite{Sarbicki2020}.

Additionally, Zhu \textit{et al}.\cite{zhu2023} derived a separability  criterion respect to correlation tensor which states that for a separable state $\rho$, the inequality as follows holds
\begin{equation}\label{zhu}
\|T_{\alpha\beta}(\rho)\|_{KF}\leq\sqrt{\|\alpha\|_2^2+\frac{d_A^2-d_A}{2}}\sqrt{\|\beta\|_2^2+\frac{d_A^2-d_A}{2}},
\end{equation}
where matrix $T_{\alpha\beta}(\rho)$ is defined by $T_{\alpha\beta}(\rho)=\left(
                                      \begin{array}{cc}
                                        \alpha\beta^t & \alpha s^t \\
                                        r \beta^t & T(\rho) \\
                                      \end{array}
                                    \right) $,
and vectors $\alpha=(a_1, ..., a_{d_A})^t$ and $\beta=(b_1, ..., b_{d_B})^t$. If we set $a_1=...=a_{d_A}=x$ and $b_1=...=b_{d_B}=y$ for Eq. (\ref{zhu}), and choose parameters $\kappa_A=\kappa_B=2$ for Eq. (\ref{mycri}), the criterion given by Ref.\cite{zhu2023} is equivalent to our criterion.

The Bloch decomposition of density matrix would change if different operator bases (with various orthogonality relations) are employed, this
also leads to different length of the Bloch vector.  When we use
the Gell-Mann operators basis $\{\lambda_i\}$,  which are the Hermitian generators of the Lie algebra $\mathfrak{su}(d)$ endowed with orthogonality relation ${ \rm{Tr}}\lambda_i\lambda_j=2\delta_{ij}$. So the orthogonal ratio $\kappa$ takes value 2.
When we take $\kappa_A=\kappa_B=2$ in Proposition \ref{prop}, we reproduce Shen's criterion \cite{shen2016}
\begin{equation}
\|\mathcal{M}_{x, y}^{(n)}\|_{tr}\leq \frac{1}{2}\sqrt{(2nx^2+d_A^2-d_A)(2ny^2+d_B^2-d_B)}.
\end{equation}

Furthermore, assigning value 1 to parameters $x$, $y$ and $n$, we get
the criterion established (Li's criterion) in Ref. \cite[Corollary 2]{liming2014}, that is,
\begin{equation}
\|\widehat{\mathcal{M}_{x, y}^{(n)}}\|_{tr}\leq
\frac{1}{2}\sqrt{(2+d_A^2-d_A)(2+d_B^2-d_B)}.
\end{equation}
While setting  variables $x=y=1$, we recover the criterion proposed by de Vicente (dV's  criteria) in Ref. \cite{vicente2007}
\begin{equation}
\|\widehat{\mathcal{M}_{x, y}^{(n)}}\|_{tr}\leq \frac{1}{2}\sqrt{(d_A^2-d_A)(d_B^2-d_B)}.
\end{equation}

Also, there are other useful traceless orthogonal operator bases besides the Gell-Mall basis, such as the Heisenberg-Weyl operator basis \cite{chang2018, zhao2020, huang2022}, defined by the following $d\times d$ matrices
\begin{equation}
W(l, m): =\sum_{k=0}^{d-1}\mathrm{exp} (\frac{2\mathrm{i}\pi kl}{d})|k\rangle\langle (k+m)\mathrm{mod} d|,
\end{equation}
where $l, m=0, 1, \cdots, d-1$,  $\{|k\rangle\}$ is the  the standard basis of the Hilbert space.
Generally, the Heisenberg-Weyl operators are unitary and non-Hermitian, and endowed with orthogonality relation ${\rm Tr} W(l, m)W(l', m')=d\delta_{l, l'}\delta_{m, m'}$, here the orthogonal ratio $\kappa$ is just equal to the dimension of the Hilbert space. If we consider the Bloch decomposition of the density matrix respect to the Heisenberg-Weyl basis, the criterion in Proposition \ref{prop} recovers the separability criteria in Ref. \cite{chang2018}, where $\kappa_A=d_A$ and $\kappa_B=d_B$.
\begin{equation}
\|\widehat{\mathcal{M}_{x, y}^{(n)}}\|\leq \sqrt{(nx^2+d_A-1)(ny^2+d_B-1)}.
\end{equation}

These discussions show that our criterion unifies most of the previous well-known criteria on entanglement.

Next we will give an example to demonstrate that our new criterion can detect more entanglement than previous ones.


%


\textbf{Example 1}
Consider the following $3\times 3$ PPT entangled state construted in \cite{ch},
\begin{equation}\label{e3}
\rho=\frac{1}{4}(I-\sum_{i=0}^4|\psi_i\rangle\langle\psi_i|),
\end{equation}
where $|\psi_0\rangle=|0\rangle(|0\rangle-|1\rangle)/\sqrt{2}$, $|\psi_1\rangle=(|0\rangle-|1\rangle)|2\rangle/\sqrt{2}$, $|\psi_2\rangle=|2\rangle(|1\rangle-|2\rangle)/\sqrt{2}$, $|\psi_3\rangle=(|1\rangle-|2\rangle)|0\rangle/\sqrt{2}$, $|\psi_4\rangle=(|0\rangle+|1\rangle+|2\rangle)(|0\rangle+|1\rangle+|2\rangle)/3$.
Let us mix $\rho$ with white noise:
\begin{equation}
\sigma_p=\frac{1-p}{9}I_9+p\rho, ~~p\in [0,1].
\end{equation}

Now we choose the Heisenberg-Weyl operators  with dimensional $d=3$ as the operator basis for Hilbert space $\mathcal{H}_A$ and $\mathcal{H}_B$. Therein the length of the Bloch vector has an upper bound $\|\textit{\textbf{v}}\|\leq \frac{\sqrt{d-1}}{d}$. Thus, the criterion inequality in Theorem \ref{thmbi1} turns into
\begin{equation}\label{ex}
\| \mathcal{M}_{x, y}^{(n)} \|_{tr}\leq \frac{1}{3d}\sqrt{({nx^2+d-1})({ny^2+d-1})}.
\end{equation}

Exactly, the entanglement threshold of state $\sigma_p$ depends on the trace norm of matrix $\mathcal{M}_{x, y}^{(n)}$. Inequality (\ref{ex}) provides an analytical expression of $\| \mathcal{M}_{x, y}^{(n)} \|_{tr}$ endowed with parameters $x$, $y$ and $n$. In order to explore the mutual influence relations between various parameters, we set $x=\frac{1}{9}y$, and analyze the inequality (\ref{ex}) in the case $n=1$, $n=2$ and $n=3$ respectively. For clarity, both sides of inequality (\ref{ex}) are drawn in the figures, here the red curved surface denotes the trace norm of $\mathcal{M}_{x, y}^{(n)}$, and the blue curved surface denotes its upper bound.
\begin{figure*}[htbp]
\begin{minipage}[t]{0.3\textwidth}
\centering
\includegraphics[width=\textwidth]{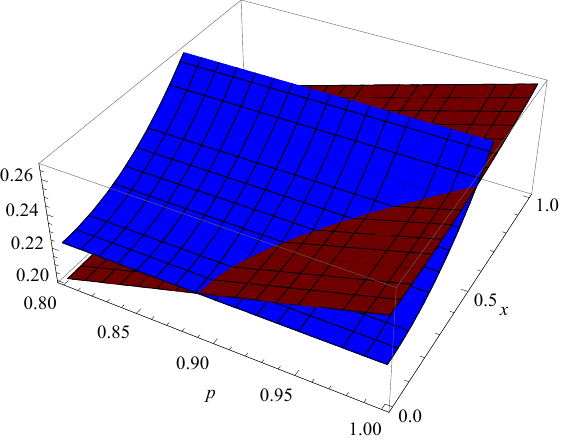}
\caption{ $n=1$.}
\end{minipage}
\begin{minipage}[t]{0.3\textwidth}
\centering
\includegraphics[width=\textwidth]{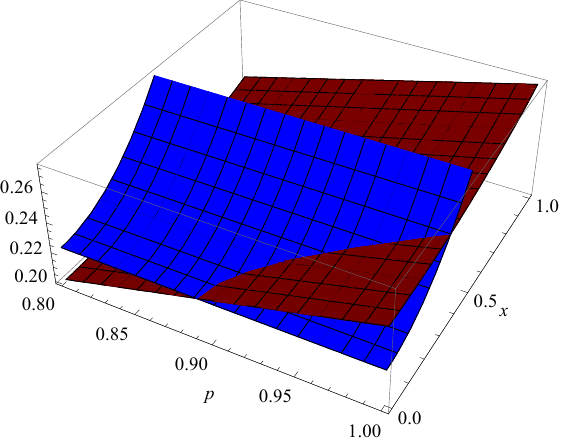}
\caption{ $n=2$.}
\end{minipage}
\begin{minipage}[t]{0.3\textwidth}
\centering
\includegraphics[width=\textwidth]{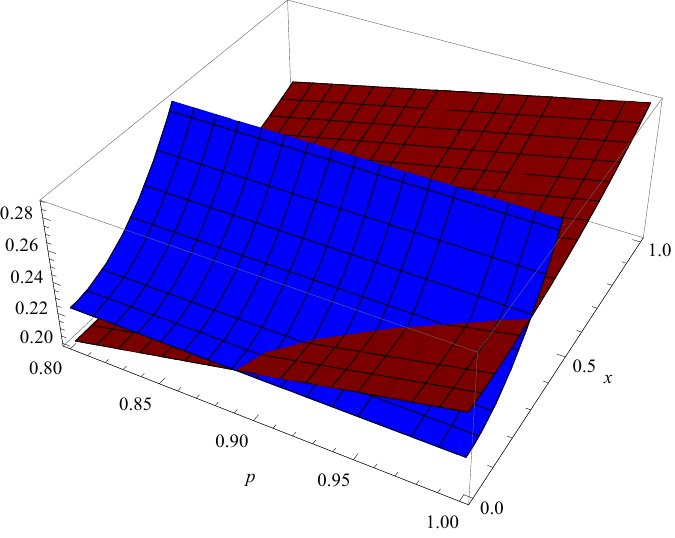}
\caption{ $n=3$.}
\end{minipage}
\end{figure*}
It can be clearly seen from the figures the critical point of entanglement  of state $\sigma_p$ is  $p\approx0.88$,  and $x$ approaches zero with $n$ increasing.

Furthermore, separability criteria such as CCNR, Li's criterion and dV's criterion can be used to detect the family of quantum states $\sigma_p$. They give rise to different detection thresholds as shown in the Table.
\begin{table}\centering
\begin{tabular}{|c| c| c| c |c |}
 \hline
 dV's criterion & CCNR  &  Li's criterion  & Theorem 1 \\ \hline
  $0.9493< p\leq 1$ & $0.8897< p\leq 1$ & $0.8925< p \leq  1$ &  $0.8843<p \leq 1$ \\ \hline
\end{tabular}
\caption{ The compare for ability of entanglement detection for various separability criteria for the state $\sigma_p=\frac{1-p}{9}I_9+p\rho$. }
\end{table}
Note that our criterion detects  entanglement for $0.8843< p \leq 1$ if we set $n=3$, $d=3$ $x=\frac{1}{27}$, $y=\frac{1}{81}$, which shows that
the new criterion is more sensitive in detecting entanglement than aforementioned separability criteria.


\textbf{Example 2}
The isotropic states in quantum system $\mathcal{H}_d\otimes \mathcal{H}_d$  are given by
\begin{equation}
\rho=\frac{1-p}{d^2}I\otimes I+p|\psi^+\rangle\langle\psi^+|,
\end{equation}
where $p\in [0, 1]$, and $|\psi^+\rangle=\frac{1}{\sqrt{d}}\sum_{i=0}^{d-1}|ii\rangle$ is the maximally entangled state. The Bloch decomposition is
\begin{equation}
\rho=\frac{1}{d^2}(I\otimes I+\frac{pd}{2}\sum_{i=1}^{\frac{(d+2)(d-1)}{2}}\lambda_i\otimes \lambda_i-\frac{pd}{2}\sum_{i=\frac{d(d+1)}{2}}^{d^2-1}\lambda_i\otimes \lambda_i).
\end{equation}
These states are known to be separable iff $p\leq \frac{1}{d+1}$ \cite{iso1999}. The extended correlation tensor $\|\widehat{\mathcal{M}_{x, y}^{(n)}}\|$ of isotropic states has singular values $\frac{xy}{d}$ and $\frac{p}{2d}$  with multiplicity $d^2-1$ for $n=2$. Now we set $x=y=\sqrt{d}$, one sees that Theorem \ref{thmbi1} implies exactly the full range of separability for the isotropic states. 
So our criterion is the strongest in this case compared with other criteria. For instance, \cite[Prop. 3]{vicente2007}
only gives the separability range as $p\leq \frac{1}{(d+1)(d-1)^2}$, and CCNR criterion just detects entanglement for $\frac{4d}{d^2-1}\leq p\leq 1$.

\textbf{Example 3}
The Werner states in bipartite system $\mathcal{H}_d\otimes \mathcal{H}_d$ are written as
\begin{equation}
\rho_W=\frac{1}{d^3-d}[(d-p)I\otimes I+(dp-1)F],
\end{equation}
where $-1\leq p\leq 1$, and $F$ is the flip operator given by $F(\phi\otimes \varphi)=\varphi\otimes \phi$. The Bloch representation of Werner states can be represented as
\begin{equation}
\rho_W=\frac{1}{d^2}(I\otimes I+\sum_i\frac{d(dp-1)}{2(d^2-1)}\lambda_i\otimes \lambda_i).
\end{equation}
These states are separable iff $p\geq 0$ \cite{werner}.
 Since singular values of extended correlation tensor $\|\widehat{\mathcal{M}_{x, y}^{(n)}}\|$ are  $\frac{xy}{d}$ and $\frac{|dp-1|}{2d(d^2-1)}$ with multiplicity $d^2-1$ for $n=2$,
employing Theorem \ref{thmbi1}, it recognizes separability for $\frac{1}{d}\leq p\leq \frac{1}{d}(\sqrt{(2x^2+d-1)(2y^2+d-1)}-2xy+1)$ and $\frac{1}{d}(2xy+1-\sqrt{(2x^2+d-1)(2y^2+d-1)})\leq p<\frac{1}{d}$. Furthermore, we set $x=y=\sqrt{d}$, thus our criterion can detect separability for $\frac{2-d}{d}\leq p\leq 1$, while the separability criterion in \cite[Prop. 3]{vicente2007} detects separability for
$ \frac{d-2}{d(d-1)}\leq p\leq \frac{1}{d-1}$. Clearly our criterion is stronger than that of \cite{vicente2007}.


\section{Application to entanglement witness}
In this section, we show that the separability criterion also gives rise to a family of entanglement witnesses. Recall that
the trace norm of any $m \times n$ matrix $X$ enjoys the following important property:
\begin{equation}\label{norm}
\|X\|_{tr}=\ \mathop{max}_{O}\mathrm{Tr}O^{\dag}X,
\end{equation}
where the maximum is over all $m\times n$ isometry matrices $O$.

In the following we fix a Hermitian operator basis for our separability criterion. Assume $\rho$ is a separable state acting on $\mathcal{H}_A\otimes \mathcal{H}_B$. Now for a fixed parametric triple $(x, y, n)$, the separability criterion in Theorem says that
\begin{equation}
 \| \mathcal{M}_{x, y}^{(n)} \|_{tr}\leq \sqrt{(\frac{nx^2 +d_A-1}{\kappa_Ad_A})(\frac{ny^2 +d_B-1}{\kappa_Bd_B})}.
 \end{equation}
 Using the property of trace norm (\ref{norm}),  we can rewrite the above expression as follows.
\begin{equation}
 \begin{aligned}
0 &\leq \sqrt{(\frac{nx^2 +d_A-1}{\kappa_Ad_A})(\frac{ny^2 +d_B-1}{\kappa_Bd_B})}-\| \mathcal{M}_{x, y}^{(n)} \|_{tr}  \\
    &=\sqrt{(\frac{nx^2 +d_A-1}{\kappa_Ad_A})(\frac{ny^2 +d_B-1}{\kappa_Bd_B})}\mathrm{Tr}(\rho I_A\otimes I_B)-\ \mathop{max}\limits_{O} \mathrm{Tr}O^{\dag} \mathcal{M}_{x, y}^{(n)}\\
  &=\sqrt{(\frac{nx^2 +d_A-1}{\kappa_Ad_A})(\frac{ny^2 +d_B-1}{\kappa_Bd_B})}\mathrm{Tr}(\rho I_A\otimes I_B)+\ \mathop{min}\limits_{O} \mathrm{Tr}O^{\dag} \mathcal{M}_{x, y}^{(n)}.
\end{aligned}
\end{equation}
where we have used the property $\mathrm{Tr}(\rho I_A\otimes I_B)=1$ and the symmetry of trace function for isometry operators, i.e., if the maximum value of $\mathrm{Tr}O^{\dag} \mathcal{M}_{x, y}^{(n)}$ can be obtained for a certain isometry $O$, simultaneously one can have the minimum value of that at $-O$.
Here both the maximum and the minimum are taking over all $(d_A^2+n-1)\times ( d_B^2+n-1)$ isometry matrices $O$, that have the same size with matrix $\mathcal{M}_{x, y}^{(n)}$.
Note that for $O=(O_{ij})$, $\mathrm{Tr}O^{\dag} \mathcal{M}_{x, y}^{(n)}=\sum_{i, j} O_{ij}m_{ij}$. Therefore we can define
a family Hermitian operator $W_O^{xy}$ of entanglement witnesses:
\begin{equation}
\mathrm{Tr}(W_{O}^{xy}\rho)\geq 0.
\end{equation}
By using the linearity of the trace function:
\begin{equation}
W_{O}^{xy}=\sum_{i, j}w_{ij}G_i^A\otimes G_j^B,
\end{equation}
where the components $w_{ij}$ are given by
\begin{equation}
\begin{split}
   & w_{00}=\frac{\sqrt{(nx^2+d_A-1)(ny^2+d_B-1)}}{\kappa_A\kappa_B}+\sum_{i=1}^{n}\sum_{j=1}^{n}xyO_{ij}, \\
   &{ w_{0,j}}=\sum_{i=1}^{d_A^2-1}xO_{i,j+n},~~ \text{for}~~ j\neq 0,~~\\
   & w_{i, 0}=\sum_{j=1}^{d_B^2-1}yO_{i+n, j}, ~~\text{for}~~ i\neq 0,\\
   & w_{i, j}=O_{i+n, j+n}, ~~\text{for}~~i\neq 0  ~~\text{and}~~ j\neq 0   .
\end{split}
\end{equation}
Here we have employed the expression of $m_{ij}$  given in \eqref{relation}.
 The entanglement witnesses $W_O^{xy}$ can be viewed parameterized by the isometry $O$.

%
%
%
%
%

\section{Conclusions}
In this paper, a feasible separability criterion is formulated in terms of the trace norm of the extended correlation tensor, which is
a parametrized criterion in detecting entanglement.
The criterion has a unified form independent from the choice of orthogonal operator bases and contain
several previously known separability criteria as special cases, including CCNR, dV's criterion and Li's criterion etc. We also use the
formulation to construct a class of
entanglement witnesses.

\bigskip
\noindent{\bf Acknowledgments}

The research is supported by the National Natural Science Foundation of China under Grant Nos. 1210301 and Nos. 12171044; the Hainan Provincial Natural
Science Foundation of China under Grant Nos.121RC539; the specific
research fund of the Innovation Platform for Academicians of Hainan
Province under Grant Nos. YSPTZX202215.

\textbf{Date Availability Statement}
All data generated or analysed during this study are available from the corresponding author on reasonable request.

\textbf{Conflict of Interest Statement}
We declare that we have no conflict of interest.

\end{document}